# Itinerant Ferromagnetism and Metamagnetism in Cr Doped Perovskite Ruthenates

V. Durairaj, E. Elhami, S. Chikara, X.N. Lin, A. Douglass and G. Cao*
Department of Physics and Astronomy
University of Kentucky, Lexington, KY40506

P. Schlottmann and E. S. Choi
National High Magnetic Field Laboratory
Florida State University, Tallahassee, FL32306

R. P. Guertin
Department of Physics and Astronomy,
Tufts University, Medford, MA 02155

We report results of structural, magnetic and transport properties of single crystal $CaRu_{1-x}Cr_xO_3$ (0≤x≤0.36) and $SrRu_{1-x}Cr_xO_3$ (0≤x≤0.30). Cr substitution as low as x=0.08 drives $CaRu_{1-x}Cr_xO_3$ from the paramagnetic state to an itinerant ferromagnetic state with field-driven first-order metamagnetic transitions leading to a sizeable saturation moment (~0.4$\mu_B$/f.u. within the ab plane). The ferromagnetism occurs abruptly and reaches as high as $T_C$=123 K for x=0.22. The Cr-driven ferromagnetism is highly anisotropic, suggesting an important role for spin-orbit coupling. Lattice constant and magnetic measurements strongly support the valence of the Cr as tetravalent ($Cr^{4+}$, $3d^2$ configuration). Cr substitution for Ru in $SrRuO_3$ ($T_C$=165 K) enhances the itinerant ferromagnetism, with $T_C$ reaching 290 K for x=0.30, consistent with Cr-induced ferromagnetism in paramagnetic $CaRuO_3$. Preliminary pressure-dependent magnetization of $CaRu_{0.85}Cr_{0.15}O_3$ shows strong enhancement of the saturation magnetization (25% for P~0.7 GPa). All results indicate a coupling of Ru 4d and Cr 3d electrons that is unexpectedly favorable for itinerant ferromagnetism which often exists delicately in the ruthenates.

The Ruddlesden-Popper (RP) series $Ca_{n+1}Ru_nO_{3n+1}$ and $Sr_{n+1}Ru_nO_{3n+1}$ (n=number of Ru-O layers/unit cell) are a class of correlated electron materials showing a rich variety of physical properties. The central characteristic of these 4d-shell based transition metal oxides is the more extended d-orbitals of the Ru-ion compared to those of 3d-shell ions. This leads to comparable and thus competing energies for the crystalline electric field (CEF) interaction, Hund's rule interactions, spin-orbit coupling, p-d hybridization and electron-lattice coupling. The deformations and relative orientations of corner-shared $RuO_6$ octahedra crucially determine the CEF level splitting and the band structure, hence the magnetic and transport properties. As a result, the physical properties are highly susceptible to perturbations such as the application of magnetic fields, pressure and slight changes in chemical composition. These features are manifested in $Ca_{n+1}Ru_nO_{3n+1}$ and $Sr_{n+1}Ru_nO_{3n+1}$: the former are on the verge of a metal-nonmetal transition and prone to antiferromagnetism whereas the latter are metallic and tend to be ferromagnetic with superconducting $Sr_2RuO_4$ (n=1) being an exception [1-32]. Such a wide variety of physical properties has not been observed in other transition metal RP systems.

The perovskites $CaRuO_3$ and $SrRuO_3$ have been extensively studied [14-36, for example] and their sharp differences in magnetic behavior are classic examples that illustrate the sensitivity of the band structure to structural distortions. Both are orthorhombic, but $SrRuO_3$ has the more "ideal" (RP, n=∞) and less distorted perovskite structure. $SrRuO_3$ is an itinerant ferromagnet with a Curie temperature $T_c$=165 K and a saturation moment $M_s$ of 1.10 $\mu_B$/Ru with the easy axis in the basal plane [19]. The CEF interaction in $Ru^{4+}$ ($4d^4$) ions is so large due to the extended 4d-orbitals that Hund's rules partially break down, yielding a low spin state with S=1 ($^3T_{1g}$). On the other hand,

CaRuO$_3$ occurs in the same crystal structure and symmetry as SrRuO$_3$, but with approximately twice as large RuO$_6$ octahedra rotation due to ionic size mismatches between Ca and Ru ions (ionic radius *r*=1.00 Å and 1.18 Å for Ca and Sr, respectively vs. r=0.620 Å for Ru$^{4+}$). This results in a ground state that is less favorable for ferromagnetism, so CaRuO$_3$ is a metallic paramagnet but verging on collective magnetism [18, 19, 34-36]. Our earlier study on single crystal Sr$_{1-x}$Ca$_x$RuO$_3$ indicates that the magnetic coupling is highly sensitive to perturbations in the Ru-O-Ru bond length and angle caused by substituting Sr with the isoelectronic but smaller Ca-ion. As a result, T$_c$ decreases monotonically with Ca concentration and vanishes for x>0.8 [19]. The sensitivity of the ground state to slight impurity doping is also evidenced in CaRu$_{1-x}$Sn$_x$O$_3$ [18], CaRu$_{1-x}$Rh$_x$O$_3$ [31], SrRu$_{1-x}$Mn$_x$O$_3$ [32], SrRu$_{1-x}$Pb$_x$O$_3$ [18] and other impurity doped CaRuO$_3$ [36] and SrRuO$_3$ [34] where substituting the impurity ion for the Ru ion extensively changes the magnetic ground state. This is very often accompanied by a metal-insulator transition.

The Cr impurity ions can form in trivalent (3d$^3$) and tetravalent (3d$^2$) states. For the Cr$^{3+}$ ion in octahedral symmetry, each of the three t$_{2g}$ orbitals is half-filled, yielding S=3/2, and for Cr$^{4+}$ ion, only two of the three t$_{2g}$ orbitals are half-filled with one orbital empty, thus S=1. The perovskites CaCrO$_3$ and SrCrO$_3$, which form only under high pressure, were found to be an insulating antiferromagnet and a metallic paramagnet, respectively [33]. The synthesis difficulty leaves many physical properties of these compounds still largely unknown. CaCrO$_3$ has the same crystal structure and symmetry as CaRuO$_3$ with a space group of *pbnm* and lattice parameters a=5.287 Å, b=5.316 Å, and c=7.486 Å [33]. This structural compatibility, as shown in Fig. 1, provides an advantage for a thorough study of $CaRu_{1-x}Cr_xO_3$ by controlling electron correlation strength without significantly altering the on-site and intersite Coulomb interaction. On the other hand, $SrCrO_3$ has a cubic perovskite structure with the *Pm3m* space group and a=3.8169 Å [33]. Recent studies on polycrystalline $SrRu_{1-x}Cr_xO_3$ show an increase in the Curie temperature $T_c$ to 188 K for x=0.11 [34, 35]. This behavior is attributed to a double-exchange interaction involving $Cr^{3+}$ [35]. In contrast, all other transition metal doping for Ru rapidly diminishes the Curie temperature. For example, $SrRu_{1-x}Mn_xO_3$ displays an evolution through a quantum critical point from itinerant ferromagnetism to insulating antiferromagnetism with increasing Mn concentration [32].

In this paper, we report an abrupt transition from paramagnetism to itinerant ferromagnetism induced by Cr doping in single crystal $CaRu_{1-x}Cr_xO_3$ for 0≤x≤0.36. The itinerant ferromagnetism, with $T_c$ as high as 123 K, occurs along with first-order metamagnetic transitions that lead to a saturation moment of 0.4 $\mu_B$/f.u. with the field aligned in the basal plane. The magnetic anisotropy in this orthorhombic system is unusually large, suggesting a critical role of spin-orbit coupling that dominates the magnetic properties. Unlike other impurity doping on the Ru site, Cr doping essentially causes no metal-insulator transition for x<0.36. As a comparison, we also present our recent results on single crystal $SrRu_{1-x}Cr_xO_3$ (0≤ x≤0.30) where $T_c$ rises from 165 to 290 K.

The single crystals of the entire series of $CaRu_{1-x}Cr_xO_3$ and $SrRu_{1-x}Cr_xO_3$ were grown using flux techniques. All crystals studied were characterized by single crystal or powder x-ray diffraction, EDS and TEM. No impurities or intergrowth were found. The magnetization was measured using the Quantum Design MPMS XL 7T magnetometer.

The resistivity was obtained using the standard four-lead technique utilizing a transport property measurement option added to the magnetometer. For each composition, a few crystals were measured and the data were averaged in order to reduce errors that could be generated by uncertain lead geometry.

Shown in Fig.1a are the lattice parameters for the a-, b- (left scale) and c-axis (right scale) as a function of Cr concentration, x, ranging from 0 to 0.36 for $CaRu_{1-x}Cr_xO_3$. The lattice parameters were determined using x-ray diffraction data on powdered crystals. For x=0 ($CaRuO_3$), the lattice parameters are in good agreement with those reported earlier [13, 19]. The orthorhombic symmetry is retained as a function of x. Within the error of the measurement, the lattice parameters generally decrease with x, consistent with the fact that the ionic radius of $Cr^{4+}$ (0.550 Å) is smaller than that of $Ru^{4+}$ (0.620 Å). The changes in the lattice parameters result in a shrinkage of the unit-cell volume by about 1.2% (x=0.36) as shown in Fig.1b. The results seem to suggest no presence of the $Cr^{3+}$ ion (0.615 Å) and/or $Ru^{5+}$ (0.565 Å), which would lead to x-dependence opposite to that shown in Fig.1. Similar changes in lattice parameters are also observed in $SrRu_{1-x}Cr_xO_3$, with the unit-cell volume reduced by about 1 % for x=0.30.

Shown in Fig. 2 is the temperature dependence of the a-axis magnetization, $M_a$, for representative compositions taken in a field cooled sequence. The major feature is the instantaneous presence of the ferromagnetic behavior upon Cr doping with a strong hysteresis effect (not shown). The Curie temperature $T_c$ increases from 70 K for x=0.08 to 115 K for x=0.15, peaks at 123 K for 0.18≤x≤0.22 and decreases to 100 K for x=0.36 (see the inset). ($T_c$ is determined as the maximum of the derivative dM/dT). The magnetic behavior is unexpectedly anisotropic as shown in Fig. 2b where the c-axis magnetization $M_c$ is much weaker than $M_a$.

The data for 200<T<350 K in Fig. 2 were fitted to a Curie-Weiss law $\chi=\chi_o+C/(T-\theta)$, where $\chi_o$ is a temperature-independent susceptibility, C is the Curie constant, and $\theta$ the Curie-Weiss temperature, for 0<x<0.36. Remarkably, the Curie-Weiss temperature $\theta_{CW}$ shows x-dependence with the same general trend of that of $T_c$, changing from -150 K through zero eventually to +120 K (see the inset in Fig.2a). The change in sign is associated with the change from a tendency to antiferromagnetic to ferromagnetic exchange coupling, consistent with the onset of ferromagnetism with increasing Cr content. The effective moment estimated from the Curie constant C decreases monotonically from 2.76 $\mu_B$/f.u. for x=0 to 1.7 $\mu_B$/f.u. for x=0.36 (see the inset in Fig.2b). These values are smaller but close to those anticipated for tetravalent Ru and Cr ions, i.e. S=1 for both $Ru^{4+}$ ($4d^4$, low spin state) and $Cr^{4+}$ ($3d^2$). Note that the effective magnetic moments for $Ru^{5+}$ (S=3/2) and $Cr^{3+}$ (S=3/2) are considerably larger, so that the Curie constant is consistent only with tetravalent Ru and Cr ions for all compositions. These results are in agreement with the structural data shown in Fig.1. Also illustrated in the inset in Fig.2b is the temperature-independent susceptibility $\chi_o$ (right scale) which stays essentially unchanged for x<0.18, but rises rapidly near x=0.18 and peaks at x=0.22 where $T_c$ reaches the maximum. $\chi_o$ is usually associated with a Pauli susceptibility and a measure of the density of states at the Fermi level, $N(\varepsilon)$, i.e., $\chi_o \sim N(\varepsilon)$. The rapid increase of $\chi_o$ may then be attributed to an increase in the density of the states.

Fig.3 shows isothermal magnetization both the a-axis and c-axis at T=2 K. The striking behavior is the metamagnetic transition with strong hysteresis for $M_a$ starting at

x=0.08 (Fig.3a). This transition then develops into a two-step transition for x=0.15 and 0.18. As shown in the inset, this two-step transition for x=0.15 depends sensitively on temperature and vanishes near 50 K. However, the magnitude of M shows only weak dependence on temperature. In addition, for the a-axis, the ordered moment, $M_s$, obtained by extrapolating M to zero-field B=0, increases initially with x from 0 for x=0 to 0.4 $\mu_B$ for x=0.18, and then decreases for x>0.18 as seen in the inset in Fig.3b. In contrast, $M_s$ for the c-axis is much smaller (see Fig.3b), suggesting an important role of the spin-orbit coupling that causes the anisotropy. It is noted that the variations of both $\chi_o$ and $M_s$ are larger in the vicinity of x=0.18, implying an intimate correlation between the density of states and the ordered moment.

Fig. 4 shows the zero field cooled T=10.5 K magnetization measured with a vibrating sample magnetometer of several randomly oriented single crystal $CaRu_{0.85}Cr_{0.15}O_3$ samples in a 13 g self-locking opposed anvil pressure cell. The data taken after cooling in zero field show the characteristic metamagnetic transition of H=2.0 T and nearly complete hysteresis. The 10.5 K M(H) was also performed at P~4 kbar and P~7 kbar and, while little shift was seen in the position of the metamagnetic transition or in the Curie temperature, a very large overall increase in M(H) was observed, attaining a 25% increase for P~7 kbar (see the inset of Fig.4). Similar results were obtained for $CaRu_{0.92}Cr_{0.08}O_3$ samples. Pressure is determined by the shift in the superconducting transition temperature of two Sn samples, one inside and one outside the clamp. The result is in contrast, for example, to the $T_c(P)$ for MnSi, a weak itinerant ferromagnet, where $T_C \sim (P_c-P)^{1/2}$, a prediction of Stoner theory [37]. Decreasing the lattice constant in ferromagnetic CaRu$_{1-x}$Cr$_x$O$_3$ may enhance the moment through band narrowing, but seems not to affect strongly T$_c$. This effect deserves further study.

Although the low temperature resistivity also undergoes significant changes with x, the metallic behavior essentially remains for all x except for x=0.36. However, as seen in Fig. 5, $\rho_{ab}$ for T<100 K rises and becomes less temperature-dependent with increasing x. Nevertheless, $\rho_{ab}$, on a logarithmic scale as a function of temperature varies by less than 3 orders of magnitude from x=0 to x=0.36 at 2 K. $\rho_{ab}$ for x=0 and 0.08 obey $\rho=\rho_o+AT^2$ ($\rho_o$ is the residual resistivity) for T<40 K. $\rho_o$ =14 and 80 $\mu\Omega$ cm and A = 45 10$^{-9}$ and 3.4 10$^{-9}$ $\Omega$ cm K$^{-2}$, for x=0 and 0.08, respectively. The increase in $\rho_o$ is expected for any impurity doping that causes more elastic scattering. The coefficient A is proportional to the square of the effective mass, m*, i.e. A~m*$^2$, and is comparable to those of other correlated electron systems. For x=0.36, $\rho_{ab}$ a slight nonmetallic behavior below 20 K and a sharp break in the slope at T$_c$ =100 K (see inset in Fig.5), which according to the Fisher-Langer theory is the consequence of scattering off short-range spin fluctuations in the neighborhood of T$_c$. We note that the Fisher-Langer behavior for other concentrations is not as strong as that for x=0.36. Finally, the negative magnetoresistance ratio at 7 T and 2K varies from 15% to 20% for x=0.15, 0.18, 0.22 and 0.36.

It is clear that low levels of substitution of Ru $t_{2g}$ electrons by Cr $t_{2g}$ electrons induces ferromagnetism and metamagnetism with strong hysteresis. In CaRuO$_3$, the 4d $t_{2g}$-orbitals are itinerant due to self-doping by the O 2p-electrons and the system is metallic. On the other hand, Cr$^{4+}$ ion based compounds have two 3d-electrons in rather contracted $t_{2g}$ orbitals, which could provide both fairly narrow band and strong exchange interactions. This is certainly true for $CrO_2$, an itinerant ferromagnet with $T_c$=450 K, where the exchange splitting between spin-up and spin-down electrons is comparable to the $t_{2g}$ bandwidth and makes 100% spin polarization possible, at least for $T \ll T_C$. The substitution of $Ru^{4+}$ by $Cr^{4+}$ replaces the 4d-electron with a more localized 3d-electron and the hybridization between the Cr 3d and Ru 4d electrons may narrow the bandwidth, W, significantly enough so that $W \sim 1/N(\varepsilon)$, motivating the occurrence of the ferromagnetism according to the Stoner model. It is also possible for ferromagnetism to readily occur if the Fermi surface $\varepsilon_F$ lies close to a sharp peak of $N(\varepsilon)$. Based on the Stoner criterion for the ferromagntism instability $U_c=1/N(\varepsilon)$ ($U_c$ is critical value of the exchange interaction between parallel-spin electrons), this could facilitate a U even smaller than W to satisfy the Stoner criterion.

The results discussed above for $CaRu_{1-x}Cr_xO_3$ are supported by a parallel study on $SrRu_{1-x}Cr_xO_3$ showing that Cr doping drastically raises the Curie temperature - from $T_C$=165 K for x=0 to $T_C$=290 K for x=0.30. This is shown in Fig. 6. The values of $T_c$ for low concentrations ($T_C$=183 K for x=0.04 and 192 K for x=0.20) are qualitatively consistent with 188 K for x=0.11 of polycrystalline samples reported earlier [34, 35]. It is new and striking that $T_c$ becomes as large as 290 K for x=0.30 though $T_c$ is largely broadened (see Fig.6). ($T_c$ is determined as the maximum of the derivative dM/dT). The field dependence of the magnetization for all x stays essentially unchanged but the saturation moment $M_s$ decreases with increasing x. $M_s$ is reduced from 1.1 $\mu_B$/f.u. for x=0 to 0.10 $\mu_B$/f.u. for x=0.30 (see the inset of Fig. 6). Nevertheless, unlike other 3d impurity doping in $SrRuO_3$, which reduce $T_c$, the Ru 4d electrons and the Cr 3d electrons are strikingly synergistic, leading to a highly enhanced exchange interaction and/or narrowed bandwidth favorable for ferromagnetism. This is particularly unusual for the perovskite ruthenates as ferromagnetism exists only in a structure that allows no significant distortions of Ru-O-Ru bond and angle [20]. Also largely unexpected is the presence of itinerant metamagnetism. Indeed, the metamagnetism may occur in a nearly ferromagnetic metal that is characterized by a maximum in magnetic susceptibility [38]. An example is $Sr_3Ru_2O_7$ which is an enhanced paramagnet with behavior consistent with proximity to a metamagnetic quantum critical point [10]. $Y(Co_{1-x}Al_x)_2$, where the Stoner ferromagnetic instability is approached by changing x, also shows a metamagnetic transition [39]. However, the robust ferromagnetic behavior and the two-step metamagnetic transition in $CaRu_{1-x}Cr_xO_3$ suggest a complex, unique band structure resulted from the 3d-4d electron coupling, though no metamagnetic transition is discerned in $SrRu_{1-x}Cr_xO_3$ (see the inset in Fig.6).

In the ruthenates transport properties, like magnetic properties, strongly depend on the relative orientation of the corner-shared octahedral, and there is a strong coupling of lattice, charge, orbital and spin degrees of freedom in general. The drastic changes in the magnetic behavior of $CaRu_{1-x}Cr_xO_3$ and $SrRu_{1-x}Cr_xO_3$ with Cr doping conspicuously accompany no metal-insulator transition which is often observed for other impurity doping [18, 31, 32, 36]. This may be associated with the fact that only two of the three Cr $t_{2g}$ levels are occupied and electron hopping between the $Cr^{4+}$ and $Ru^{4+}$ $t_{2g}$ orbitals is energetically favorable, so the dynamic itinerant character of the d-electrons is retained. The impurity doping, of course, introduces defects and disorder raising the electrical resistivity at low temperatures and this less metallic behavior for large x could be also associated with a site percolation of nearest neighbor Ru-Ru bonds [40]. The disruption of Ru connectivity affects the orientation of the RuO$_6$ octahedra (tilting angle), which to a great extent determines the properties of the ruthenates.

Unlike all other impurity doping for the Ru site, slight Cr-doping facilitates the presence of the ferromagnetism that is extremely delicate in the perovskite ruthenates. The same effect is also seen in the triple layered Sr$_4$Ru$_3$O$_{10}$ [41]. Apparently, the Ru 4d-electrons and Cr 3d-electrons are unusually synergistic to promote ferromagnetism in these materials. The phenomena merit additional experimental and theoretical investigation.

**Acknowledgements**

G.C. is indebted to Dr. Joseph Budnick for valuable discussions that resulted in this work. This work was supported by National Science Foundation under grant No. DMR-0240813. P.S. acknowledges the support by NSF (grant No. DMR01-05431) and DOE (grant No. DE-FG02-98ER45707). R. P. G. acknowledges the support of NHMFL VSP #43.

*Corresponding author. Email:cao@uky.edu

Captions:

**Fig.1.** (a) Lattice parameters for the a-, b- (left scale) and c-axis (right scale) and (b) The volume of the unit V as a function of Cr concentration x for $CaRu_{1-x}Cr_xO_3$.

**Fig.2.** (a) Temperature dependence of the a-axis magnetization M for representative compositions for $CaRu_{1-x}Cr_xO_3$; Inset: the Curie temperature Tc (left scale) and the Curie-Weiss temperature $\theta_{CW}$ (right scale) as a function of x. (b) Temperature dependence of the c-axis magnetization M for a few representative x; Inset: The effective moment $\mu_{eff}$ (left scale) and temperature-independent susceptibility $\chi_o$ (right scale) as a function of x.

**Fig.3.** (a) Isothermal magnetization M for the a-axis at T=2 K $CaRu_{1-x}Cr_xO_3$; Inset: M vs B for x=0.15 for various temperatures. (b) Isothermal magnetization M for the c-axis at T=2 K; Inset: the saturation moment $M_s$ for the a- and c-axis direction.

**Fig.4.** Isothermal magnetization M $CaRu_{0.85}Cr_{0.15}O_3$ at T=10.5 K and P=7 kbar measured in a zero field cooled sequence using a vibrating sample magnetometer of several randomly oriented single crystal $CaRu_{0.85}Cr_{0.15}O_3$ samples in a 13 g self-locking opposed anvil pressure cell; Inset: M as a function of P.

**Fig.5.** (a) Basal plane resistivity $\rho_{ab}$ on a logarithmic scale vs. temperature for x=0, 0.08, 0.15, 0.18 and 0.36; Inset: $\rho_{ab}$ (left scale) and M (right scale) as a function of temperature for x=0.36.

**Fig.6.** Temperature dependence of the a-axis magnetization M for $SrRu_{1-x}Cr_xO_3$ for x=0, 0.04, 0.20 and 0.30; Inset: Isothermal magnetization M for the ab-plane at T=2 K.

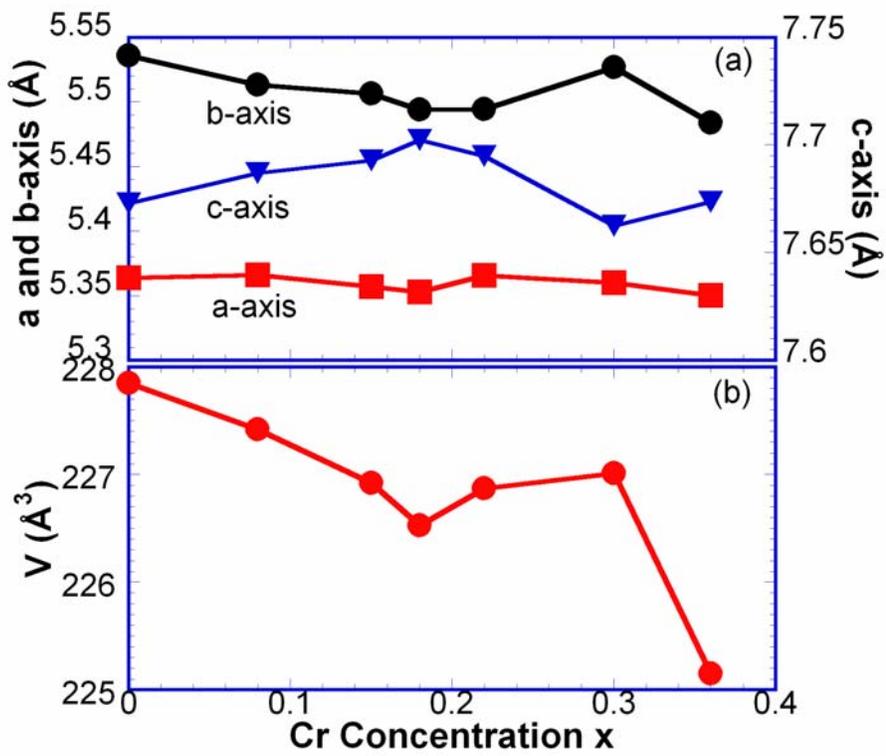

Fig. 1

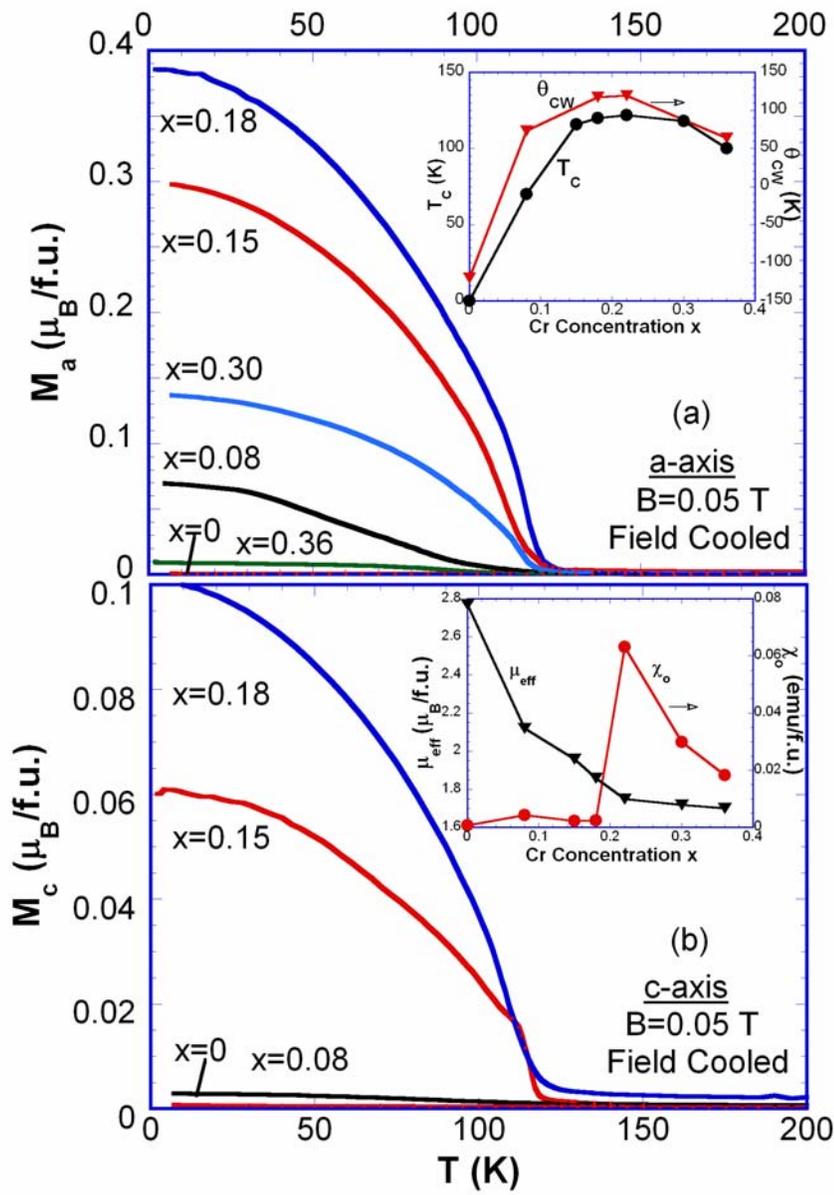

Fig. 2

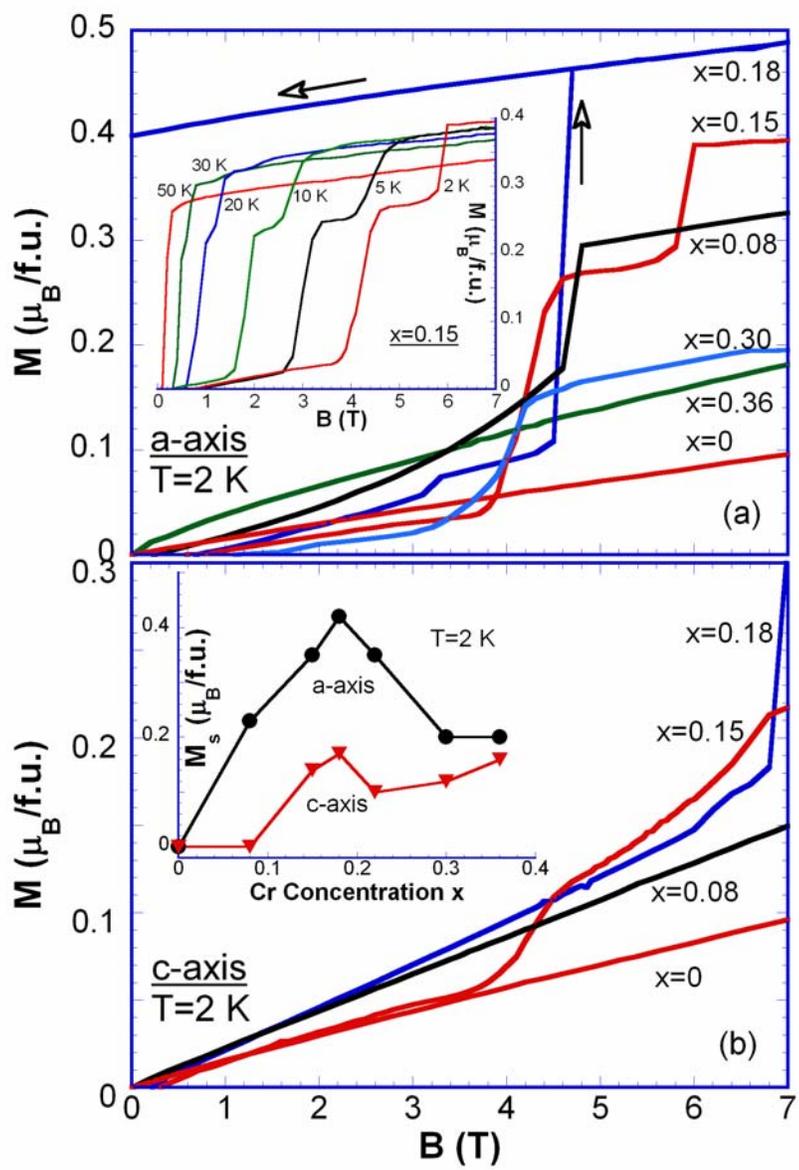

Fig. 3

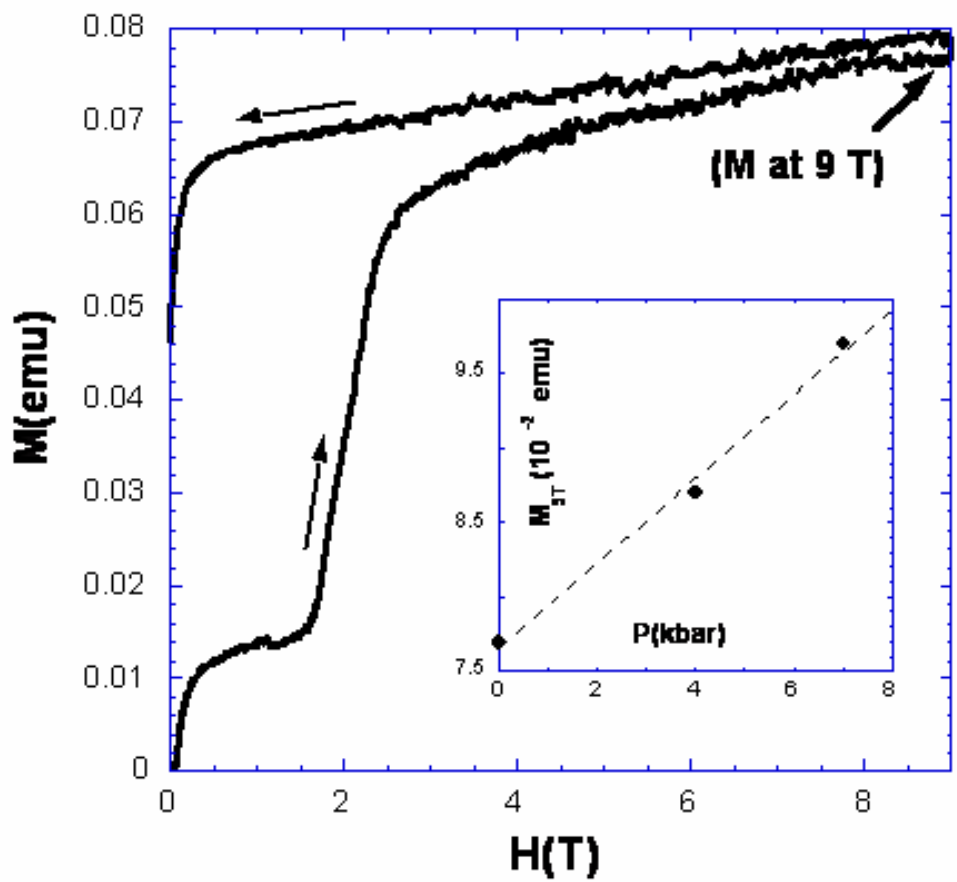

Fig. 4

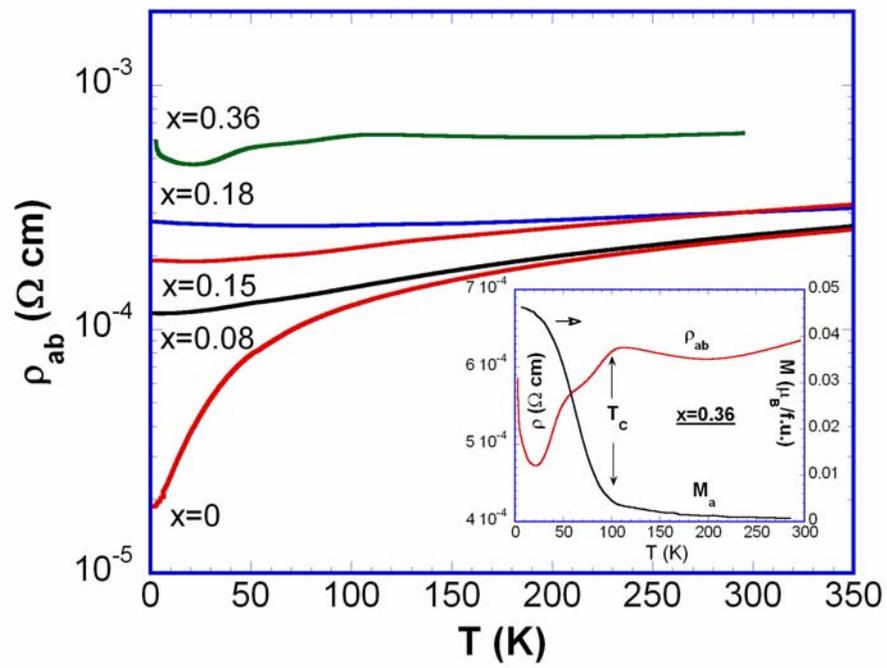

Fig.5

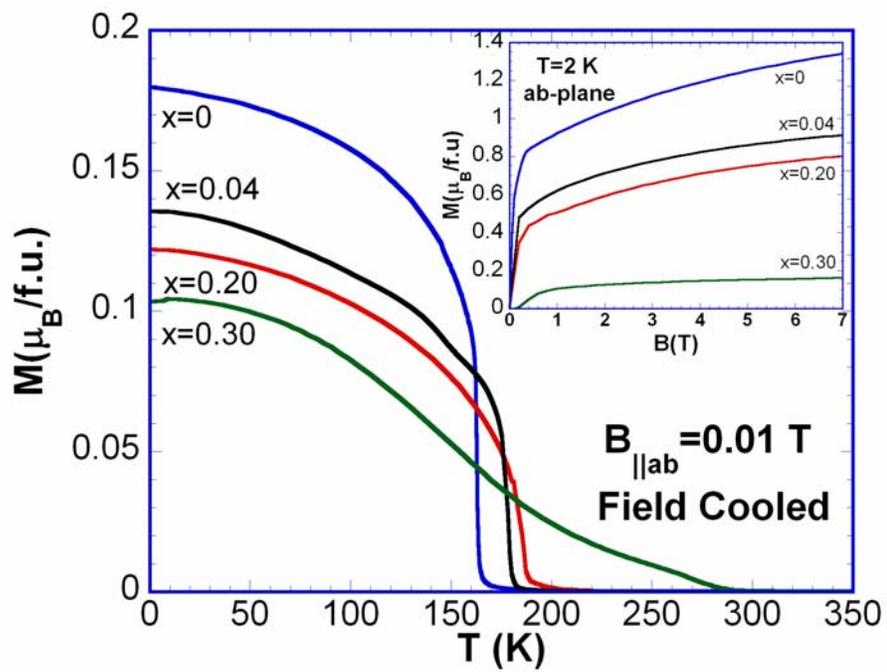

Fig. 6